\documentclass{elsart}
\usepackage{amssymb}
\usepackage{amsmath}
\usepackage[dvips]{graphicx}

\journal{:   June 29, 1987, (in French
version).\qquad\qquad\qquad\qquad\qquad\qquad\qquad\qquad\qquad\qquad\qquad\qquad\qquad\qquad\qquad\qquad\qquad\qquad\qquad
}
\input{tcilatex}
\begin{document}

\begin{frontmatter}

\title{Thermodynamic form of the equation of motion for perfect fluids of
grade $n$
 \\
  \textbf{\footnotesize\emph{ Translation    of \
C. R. Acad. Sci. Paris t. 305, II, p. 833-838 (1987)}}}

\author{Henri Gouin}
\ead{henri.gouin@univ-cezanne.fr}

\address {
 Universit\'e d'Aix-Marseille \&   C.N.R.S.  U.M.R. 6181, \\ Case 322, Av. Escadrille
 Normandie-Niemen, 13397 Marseille Cedex 20 France}

\begin{abstract}
We propose a thermodynamic form of the equation of motion for
perfect fluids of grade $n$ which  generalizes the one given by J. Serrin in
the case of perfectly compressible fluids (\cite{1}, p. 171). First integrals and
circulation theorems are deduced and a classification of the flows is given.
\end{abstract}


\end{frontmatter}

\section{Introduction}

In continuum mechanics, first gradient media cannot give a model for fluids
with strong variations of density. Material surfaces need their own
characteristic behavior and properties of energy \cite{2}.\ D.J.\ Korteweg
has been the first to point out the convenience of fluids of grade upper
than one and not to model interfacial layers by means of discontinuity
surfaces in liquid-vapor interfaces \cite{3}. Recently, this approach was
used in the case of dynamic changes of phases \cite{4}.

Till now, the thermodynamics of fluids was neglected and it was not possible
to model flows with strong variations of temperature such as those
associated with combustion phenomena or non isothermal interfaces. In fact,
it is difficult to take the gradients of temperature into account (it is not
possible to consider a virtual displacement of temperature) but we can use
the specific entropy through internal energy density.

To this aim, we usefully describe conservative flows of a compressible fluid
by the means of its internal energy depending on both entropy and density.
This will only be a limit mathematical model, however the study of
mathematical structure of the equations of motion is fundamentally
necessary. So, to improve the model for strong variations of entropy and
density in interfacial layers we consider an internal energy function of the
two quantities and their spatial gradients up to a $n-1$ convenient order:
the fluid will be said \textit{of grade }$n$ \cite{5}.

The paper aims to show that the equations of motion for perfect fluids of
any grade can be written in an universal thermodynamic form structurally
similar to the one given by J. Serrin in the case of conservative perfect
fluids \cite{1}. When the thermodynamic form is applied to the second
gradient fluids \cite{6,7}, it leads to three results: first integrals
associated with circulation theorems (such as Kelvin theorems), potential
equations representing the motion of the fluid \cite{8,9} and a classification
of the flows similar to the one of compressible perfect fluids \cite{10,11}.

\section{Fluids of grade $n$.}

Perfect fluids of grade $n$ ($n$ is any integer) are continuous media with
an internal energy per unit mass $\varepsilon$ which is a function of the
specific entropy $s$ and the density $\rho $ in the form : %
\begin{equation*}
\varepsilon =\varepsilon (s,\func{grad}s,...,(\func{grad})^{n-1}s,\rho ,%
\func{grad}\rho ,...,(\func{grad})^{n-1}\rho )
\end{equation*}%
where $(\func{grad})^{p}$, $p\in \{1,...,n-1\}$, denotes the successive
gradients in the space $D_{t}$ occupied by the fluid at present time. We
easily may consider the case of an inhomogeneous fluid but, for the sake of
simplicity  we will not do it. So, the material is supposed to have
infinitely short memory and the motion history until an arbitrarily chosen
past does not affect the determination of the stresses at present time.

\section{Equation of motion written in thermodynamic form.}

The virtual works principle (or the virtual powers principle) is a
convenient way to find the equation of motion. For conservative motions, it
writes as the Hamilton principle \cite{8}. \newline
A particle is identified in Lagrange representation by the position $\mathbf{%
X}(X_{1},X_{2},X_{3})$ occupied in the reference space ${D}_{0}$. At time $t,
$ its position is given in ${D}_{t}$ by the Eulerian representation $\mathbf{%
x}(x_{1},x_{2},x_{3}).$

The variations of particles motion are deduced from the function family :
\begin{equation}
\mathbf{X}=\mathbf{\psi }(\mathbf{x},t;\alpha )
\end{equation}%
where $\alpha $ denotes the parameter defined in the vicinity of $0$
associated with a family of virtual motions of the fluid. The real motion
corresponds with $\alpha =0$ \cite{1}.

Virtual displacements associated with any variation of the real motion can
be written in the form :
\begin{equation*}
\delta \mathbf{X}=\left. \frac{\partial \mathbf{\psi }}{\partial \alpha }
(\mathbf{x},t;\alpha )\right\vert _{\alpha =0}.
\end{equation*}%

This variation is \textit{dual} and mathematically equivalent to Serrin's
one (\cite{1}, p. 145, \cite{8}). Let $L$ be the Lagrangian of the fluid of
grade $n$ :
\begin{equation*}
L=\rho \left(\frac{1}{2}\mathbf{V}^{\ast }\mathbf{V}-\varepsilon -\Omega
\right)
\end{equation*}%
where $\mathbf{V}$ denotes the velocity of particles, $\Omega $ the
potential of mass forces defined on ${D}_{t}$ and $^{\ast }$ the
transposition in ${D}_{t}$. Between times $t_{1}$ and $t_{2} $, the \emph{%
Hamilton action} writes \cite{1,8} :%
\begin{equation*}
a=\int_{t_{1}}^{t_{2}}\int_{{D}_{t}}L~dv~dt.
\end{equation*}
where $dv$ denotes the volume element.

The density satisfies the conservation of mass :
\begin{equation}
\rho \,\det \mathbf{F}=\rho _{0}(\mathbf{X})
\end{equation}%
where $\rho _{0}$ is defined on ${D}_{0}$ and $\mathbf{F}$ is the gradient
of deformation. The motion is supposed to be conservative, then the specific
entropy is constant along each trajectory :
\begin{equation}
s=s_{0}(\mathbf{X}).
\end{equation}
Classical calculus of variations yields the variation of Hamilton action :

From
\begin{equation*}
\delta a=a^{\prime }(\alpha )_{|{\alpha =0}},
\end{equation*}%
we deduce,
\begin{eqnarray}
\delta a &=&\int_{t_{1}}^{t_{2}}\int_{{D}_{t}}\Big[(\frac{L}{\rho }-\rho\,
\varepsilon _{\rho }^{\prime })\delta \rho +\rho\, V_{i}\delta V_{i}-\rho\,
\varepsilon _{s}^{\prime }\delta s \\
&&\qquad\quad -\rho\, (\varepsilon ,_{\rho ,_{i}}\delta \rho
,_{i}+...+\varepsilon ,_{\rho ,_{i_{1}...i_{n-1}}}\delta \rho
,_{i_{1}...i_{n-1}}+\varepsilon ,_{s,_{i}}\delta s,_{i}+ \ldots  \notag \\
&&\qquad\qquad\qquad\qquad\qquad\quad +\, \varepsilon
,_{s,_{i_{1}...i_{n-1}}}\delta s,_{i_{1}...i_{n-1}})\Big] %
~dx_{1}~dx_{2}~dx_{3}~dt.  \notag
\end{eqnarray}%
The definition of \emph{dual} virtual motions yields :%
\begin{equation*}
\begin{array}{ccccc}
\delta (\func{grad})^{p}\rho =(\func{grad})^{p}\delta \rho &  & \text{and} &
& \delta (\func{grad})^{p}s=(\func{grad})^{p}\delta s.%
\end{array}%
\end{equation*}
By using the Stokes formula, let us integrate by parts. Virtual
displacements are supposed to be null in the vicinity of the edge of ${D}_{t}
$ and integrated terms are null on the edge. We deduce :%
\begin{eqnarray}
\delta a &=&\int_{t_{1}}^{t_{2}}\int_{{D}_{t}}\Big\{ \Big[ (\frac{L}{\rho }%
-\rho\, \varepsilon _{\rho }^{\prime }-\sum_{p=1}^{n-1}(-1)^{p}(\rho\,
\varepsilon ,_{\rho ,_{i_{1}...i_{p}}}),_{i_{1}...i_{p}}\Big] \delta \rho \\
&&\qquad- \Big[ \rho\, \varepsilon _{s}^{\prime
}+\sum_{p=1}^{n-1}(-1)^{p}(\rho\, \varepsilon
,_{s,_{i_{1}...i_{p}}}),_{i_{1}...i_{p}}\Big] \delta s-\rho\, V_{i}\delta
V_{i}\Big\} ~dx_{1} dx_{2} dx_{3}~dt.  \notag
\end{eqnarray}
With $\func{div}_{p}$ denoting the divergence operator iterated $p$ times on
the edge of ${D}_{t}$, we obtain Rel. (5) in tensorial form :%
\begin{eqnarray*}
\delta a &=&\int_{t_{1}}^{t_{2}}\int_{{D}_{t}}\left\{ \left[ \frac{L}{\rho }%
-\rho\, \varepsilon _{\rho }^{\prime }-\sum_{p=1}^{n-1}(-1)^{p}\,{\func{div}}%
_{p}\left( \rho\, \frac{\partial \varepsilon }{\partial (\func{grad}%
)^{p}\rho }\right) \right] \right. ~\delta \rho \\
&&\qquad\quad -\left. \left[ \rho \, \varepsilon _{s}^{\prime
}+\sum_{p=1}^{n-1}(-1)^{p}\,{\func{div}}_{p}\left( \rho\, \frac{\partial
\varepsilon }{\partial (\func{grad})^{p}s}\right) \right] \delta s-\rho\,
\mathbf{V}^{\ast }\delta \mathbf{V}\right\} ~dv~dt .
\end{eqnarray*}

By taking (2) into account, we obtain :
\begin{equation*}
\delta \rho =\rho\, {\func{div}}_{0}\,\delta \mathbf{X}+\frac{1}{\det
\mathbf{F}}\frac{\partial \rho _{0}}{\partial \mathbf{X}}\,\delta \mathbf{X}
\end{equation*}%
where $\func{div}_{0}$ denotes the divergence operator relatively to
Lagrange variables in ${D}_{0}.$

We also get :%
\begin{equation*}
\delta s=\frac{\partial s_{0}}{\partial \mathbf{X}}\,\delta \mathbf{X.}
\end{equation*}%
The definition of velocity implies :%
\begin{equation*}
\frac{\partial \mathbf{X}}{\partial \mathbf{x}}(\mathbf{x},t)\mathbf{V}+%
\frac{\partial \mathbf{X}}{\partial t}(\mathbf{x},t)=0,
\end{equation*}%
therefore%
\begin{equation*}
\frac{\partial \delta \mathbf{X}}{\partial \mathbf{x}}\mathbf{V}+\frac{%
\partial \mathbf{X}}{\partial \mathbf{x}}\delta \mathbf{V}+\frac{\partial
\delta \mathbf{X}}{\partial t}=0.
\end{equation*}%
Let us consider
\begin{equation*}
\delta \mathbf{V}=-F\overset{^{\huge \centerdot }}{\delta \mathbf{X}},
\end{equation*}%

where $\overset{\,\Huge \,^{^\centerdot }}{}\,$ denotes the material derivative. Denoting
\begin{equation*}
\begin{array}[t]{l}
\displaystyle p=\rho ^{2}\varepsilon _{\rho }^{\prime }+\rho
\sum_{p=1}^{n-1}(-1)^{p}{\func{div}}_{p}\left( \rho\ \frac{\partial
\varepsilon }{\partial (\func{grad})^{p}\rho }\right)  \\
\\
\displaystyle\theta =\varepsilon _{s}^{\prime }+\frac{1}{\rho }%
\sum_{p=1}^{n-1}(-1)^{p}{\func{div}}_{p}\left( \rho\ \frac{\partial
\varepsilon }{\partial (\func{grad})^{p}s}\right)  \\
\\
\begin{array}[t]{lllll}
\displaystyle h=\varepsilon +\frac{p}{\rho } &  & \text{and} &  & %
\displaystyle m=\frac{1}{2}\mathbf{V}^{\ast }\mathbf{V}-h-\Omega,
\end{array}%
\end{array}%
\end{equation*}%
then, Rel. (5) yields :
\begin{eqnarray*}
\delta a &=&\int_{t_{1}}^{t_{2}}\int_{{D}_{t}}\left[ m~\delta \rho -\rho
\,\theta ~\delta s+\rho \,\overset{\centerdot }{(\mathbf{V}^{\ast }\mathbf{F}%
)}\delta \mathbf{X}\right] ~dv~dt \\
& =& \int_{t_{1}}^{t_{2}}\int_{{D}_{0}}\rho _{0}\left[ (%
\overset{\centerdot }{\mathbf{V}^{\ast }\mathbf{F})}-\theta \,{\func{grad}}%
_{0}^{\ast }s-{\func{grad}}_{0}^{\ast }m\right] ~\delta \mathbf{X}~dv_{0}~dt
\end{eqnarray*}%
where $\func{grad}_{0}$ denotes the gradient operator in ${D}_{0}.$

The principle \textit{for any displacement }$\delta \mathbf{X}$\textit{\
null on the edge of }$D_{0}$\textit{, }$\delta a=0 $ implies:%
\begin{equation}
\overset{\centerdot }{\mathbf{V}^{\ast }\mathbf{F}}=\theta\, {\func{grad}}%
_{0}^{\ast }\,s_{{0}}+{\func{grad}}_{0}^{\ast }\,m.
\end{equation}%
Noting that $(\mathbf{\Gamma }^{\ast }+\displaystyle\mathbf{V}^{\ast }\frac {%
\partial \mathbf{V}}{\partial \mathbf{x}})\mathbf{F}=\overset{\centerdot }{%
\mathbf{V}^{\ast }\mathbf{F}}$, we get :%
\begin{equation}
\mathbf{\Gamma }=\theta \func{grad}s-\func{grad}(h+\Omega).
\end{equation}%
Taking Rel. (3) into account, we obtain :%
\begin{equation*}
\overset{\centerdot }{s}=0.
\end{equation*}%
Relation (7) is the generalization of Rel. (29.8) in \cite{1}. This is a
thermodynamic form of the equation of motion of perfect fluids of grad $n$.

Obviously, term $p$ has the same dimension as   pression, $\theta $ has the
same dimension as   temperature and $h$ has the same dimension as   specific
enthalpy. It seems natural to call them \textit{pression, temperature and
enthalpy of the fluid of grad }$n$, respectively.

\section{Conservative properties of perfect fluids of grad $n$.}

Relation (7) leads to the same conclusions as those obtained in \cite{1},
\cite{7,8,9,10,11} but for fluids of grad $n$. Let us remind the most important
results.

With $J$ denoting the circulation of velocity vector along a closed fluid
curve $\mathcal{C}$ convected by the flow,
\begin{equation*}
\frac{dJ}{dt}=\int_{\mathcal{C}}\theta ~ds.
\end{equation*}

The\textit{\ Kelvin theorems} are deduced: the circulation of the velocity
vector along a closed, isentropic (or isothermal) fluid curve is constant.

For any motion of fluids of grad $n$, we can write the velocity field in the form:
\begin{equation}
\mathbf{V}=\func{grad}\varphi +\psi \func{grad}s+\tau \func{grad}\chi ,
\end{equation}
the scalar potentials $\varphi, \psi, s, \tau $ and $\chi$ verifying :%
\begin{equation}
\begin{array}[t]{lllllllll}
\overset{\centerdot }{\varphi }=\displaystyle\frac{1}{2}\mathbf{V}^{\ast }\mathbf{V}%
-h-\Omega , &  & \overset{\centerdot }{\tau }=0, &  & \overset{\centerdot }{%
\psi }=\theta , &  & \overset{\centerdot }{\chi }=0, &  & \overset{%
\centerdot }{s}=0 .
\end{array}%
\end{equation}%
 Eqations (8) and (9) induce the same classification as for conservative
flows of compressible perfect fluids \cite{8,9} :

\bigskip

\textit{Oligotropic motions. - } They are motions for which surfaces of
equal entropy are vortex surfaces. The circulation of the velocity vector
along a closed, isentropic fluid curve is null. Equation (8) of the motion
yields :
\begin{equation*}
\mathbf{V}=\func{grad}\varphi +\psi \func{grad}s.
\end{equation*}

\bigskip

\textit{Homentropic motions}. - In the whole fluid $s$ is constant and Eq.
(8) yields :
\begin{equation*}
\mathbf{V}=\func{grad}\varphi +\tau \func{grad}\chi \, .
\end{equation*}%
The Cauchy theorem can be easily written :%
\begin{equation*}
\frac{d}{dt}\,(\frac{\func{rot}\mathbf{V}}{\rho })=\frac{\partial \mathbf{V}%
}{\partial \mathbf{x}}\frac{\func{rot}\mathbf{V}}{\rho }\,.
\end{equation*}%
Denoting $\mathcal{H}=\frac{1}{2}\mathbf{V}^{\ast }\mathbf{V}+h+\Omega ,$
Eq. (7) yields the Crocco-Vazsonyi equation generalized to stationary
motions of perfect fluids of grad $n$ :
\begin{equation*}
\func{rot}\mathbf{V}\times \mathbf{V}=\theta \func{grad}s-\func{grad}%
\mathcal{H}.
\end{equation*}

The laws of conservation expressed by the Kelvin theorems correspond to the
group of permutations of particles of equal entropy.
\\
This group keeps the equations of motion invariant. It is associated to an
expression of Noether's theorem as in \cite{12}. So, it is natural to
conjecture such results for a general fluid whose internal energy is a
functional of the density and the entropy.

\end{document}